\documentclass[aps,reprint,twocolumn,superscriptaddress,amsmath,amssymb]{revtex4-1}
\usepackage{color}
\usepackage{graphicx}
\usepackage[colorlinks=true, urlcolor=blue, linkcolor=blue, citecolor=blue]{hyperref}
\usepackage{ulem}
\usepackage{footnote}

\usepackage{subfigure}
\usepackage{xcolor}

\begin{document}

\title{Topological Surface States in Thick Partially Relaxed HgTe Film}

\author{M.\,L.\,Savchenko}
\author{D.\,A.\,Kozlov}
\author{N.\,N.\,Vasilev}
\author{Z.\,D.\,Kvon}
\author{N.\,N.\,Mikhailov}
\affiliation{Novosibirsk State University, Novosibirsk 630090, Russia}
\affiliation{Rzhanov Institute of Semiconductor Physics, Novosibirsk 630090, Russia}

\author{S.\,A.\,Dvoretsky}
\author{A.\,V.\,Kolesnikov}
\affiliation{Rzhanov Institute of Semiconductor Physics, Novosibirsk 630090, Russia}
\date{\today}

\begin{abstract}
Surface states of topological insulators (TIs) have been playing the central role in the majority of outstanding investigations in low-dimensional electron systems for more than 10 years. 
TIs based on high quality strained HgTe films demonstrate a variety of subtle physical effects. 
The strain leads to a bulk band gap, but limits a maximum HgTe strained film thickness, and, therefore, the majority of experiments were performed on the films with a thickness of less than 100\,nm. 
Since a spatial separation of topological states is crucial for the study of a single surface response, the HgTe thickness is essential to be increased further. 
In this work, by combining transport measurements together with capacitance spectroscopy, we performed the analysis of a 200\,nm partially relaxed HgTe film. 
The Drude fit of the classical magnetotransport revealed the ambipolar electron-hole transport with a high electron mobility. 
The detailed analysis of Shubnikov-de Haas oscillations, both in conductivity and capacitance, allowed us to distinguish three groups of electrons, identified as electrons on top and bottom surfaces and bulk electrons. 
The indirect bulk energy gap value was found to be close to zero. 
It was established that the significant gap decrease does not affect the surface states which are found to be well-resolved and spin non-degenerate. 
The presented techniques allow the investigations of other 3D TIs, regardless of the presence of bulk conductivity.
\end{abstract}

\maketitle

\section{Introduction}
An ideal three-dimensional topological insulator (3D TI) is a system that has spin non-degenerate surface states with spin-momentum locking and insulating bulk \cite{Ando2013}.
The unique properties of 3D TIs result in many new physical effects: spin polarization of 3D TI current \cite{Fan2016}, $4\pi$ oscillations in a superconductor contact \cite{Wiedenmann2016}, a universal magnetoelectric effect \cite{Dziom2017a}, Majorana fermions \cite{Sato2017}.
The existence of the mentioned surface states was established in Bi-, Sb-based and HgTe 3D TIs by means of ARPES \cite{Xia2009-2,Pan2011, Crauste2013, Brune2011}.
Then they were studied in magneto-optic \cite{Dziom2017a, Dantscher2015, Hanson2016} and magnetotransport experiments \cite{Culcer2012, Brahlek2014, Brune2011, Kozlov2014, Maier2017}, and by magnetocapacitance spectroscopy \cite{Kozlov2016}.

There are several issues hampering the investigation of the surface states. The main shortcoming  of Bi- and Sb-based 3D TIs is a crystalline imperfection leading to intrinsic doping and reduced carriers mobility. 
Only recently, the fabrication of samples with pure surface conductivity was difficult to implement and one had to study either a transport response of the surface states intermixed with the bulk ones
\cite{Chen2010, Kim2012, Bao2012, Wolgast2013}, or to use other techniques, e.g.\ studying photogalvanic effects involving a space symmetry analysis \cite{Olbrich2014}, which allowed distinguishing the bulk and surface carriers. 
Despite, in recent papers, the surface-dominant conductivity was conclusively shown \cite{Tian2014, Xu2014, Yoshimi2015,Li2017}, the best samples still have significant disorder, smearing fine effects: for instance, a pronounced QHE plateaux requires a magnetic field close to the limit of superconductive magnets or even exceeding it \cite{Xu2014, Yoshimi2015, Li2017}.

In contrast, strained HgTe-based 3D TIs are characterized by much smaller disorder and higher electron mobility values exceeding 10$^5$\,cm$^2$/V$\cdot$s \cite{Kozlov2014}. 
Moreover, the highest quality of such systems allowed studying delicate ballistic and interference effects \cite{Maier2017, Ziegler2018}. However, these systems have their own shortcomings mainly coming from a zero energy gap in bulk HgTe. 
In order to open the gap and organize a 3D TI, one has to apply a strain: HgTe films, grown on a CdTe substrate with a 0.3\%
lattice mismatch, have the indirect band gap of 15\,meV. 
In turn, the requirement of the strain limits the maximum HgTe film  thickness to about 100-150\,nm because of its relaxation at higher thicknesses (according to \cite{Brune2011} and our experience). 
This fact explains why in the majority of papers devoted to the HgTe-based 3D TIs only 70-100\,nm films were studied \cite{Brune2011, Shuvaev2012, Shuvaev2013a, Kozlov2014, Wiedenmann2016, Wiedenmann2017, Maier2017, Ziegler2018}. 
On the other hand, studying thicker HgTe films is also desirable because of the higher separation between the surface states, weaker mutual electrostatic coupling and possible hybridization. 
One could expect that the thicker films would  be partially or fully  relaxed resulting in a smaller or even zero effective energy gap. 
In such system the bulk carriers would inevitably appear at all Fermi level positions.
Based on the foregoing, the issue of a surface and bulk states coexistence seems to be actual. 
Moreover, there are some materials predicted to have a non-trivial topology with a lack of a band gap or with band overlapping \cite{Muchler2012, Bradlyn2017}.

In this paper, we investigate 200\,nm HgTe films that can be characterized by the partial strain. 
The samples under investigation are equipped with a metallic top gate allowing to widely vary a Fermi level position from the valence to conduction bands. 
By combining classical and quantum magnetotransport studies, as well as magnetocapacitance spectroscopy, we show that both a significant decrease of the bulk gap and the presence of trivial bulk conductivity do not affect the surface states.
The detailed analysis reveals the main properties of the surface and bulk carriers and proves that the surface states are still spin non-degenerate.

\section{Experimental details}\label{sec: Experimental details}
All measurements are carried out on 200\,nm HgTe films that have been grown by molecular beam epitaxy on a GaAs(013) substrate with the same layer ordering as it was for usual 80\,nm films \cite{Kozlov2014}.

\begin{figure}
    \includegraphics[width=1\columnwidth]{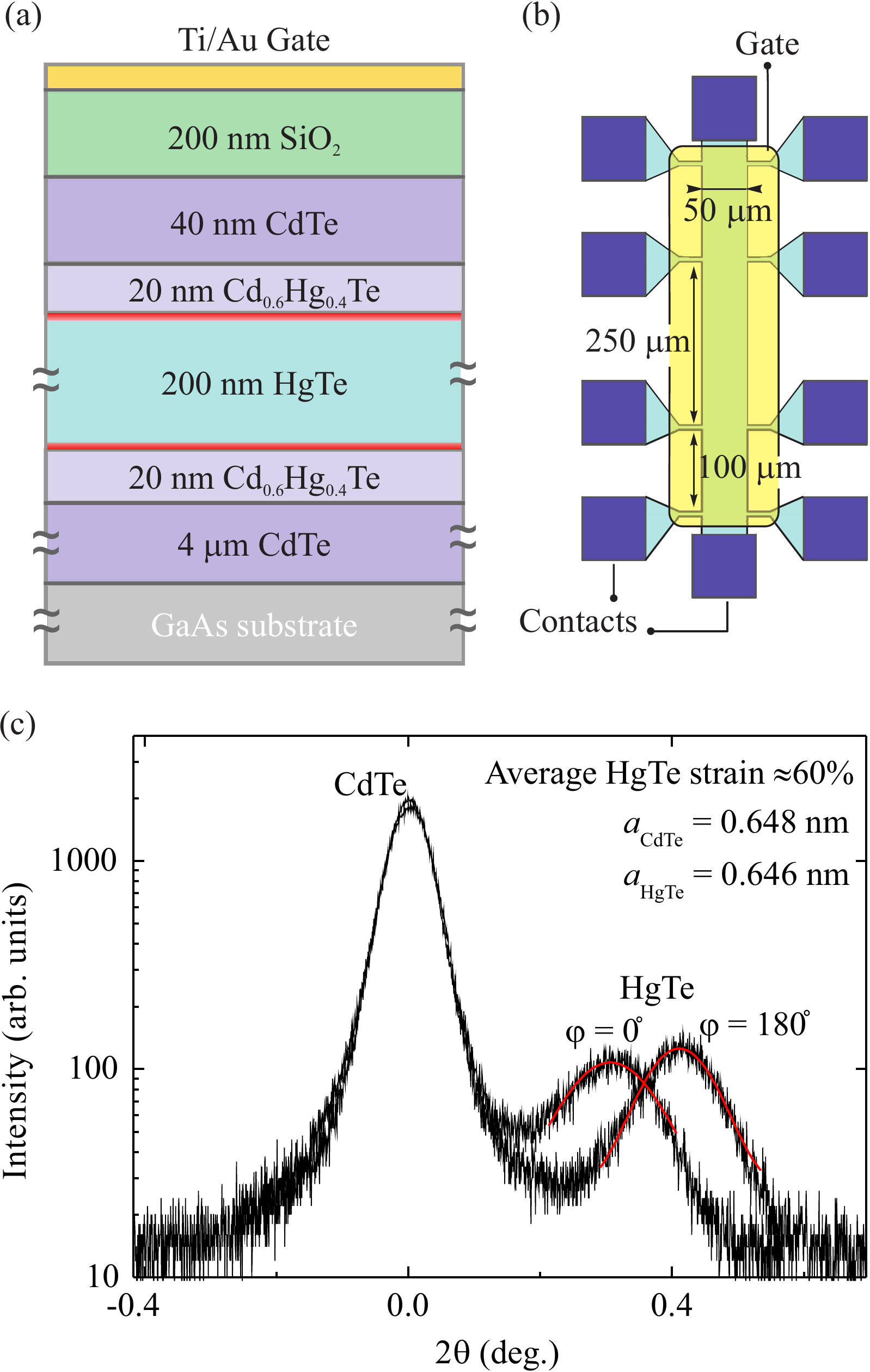}
   \caption{
   (a)  Schematic cross-section of the structure under study.
   The 200\,nm HgTe film is sandwiched between thin Cd$_{0.6}$Hg$_{0.4}$Te buffer layers covered by a 200\,nm SiO$_2$ insulator and a metallic gate.
   Bright red lines represent the surface states on the top and bottom surfaces of the film.
   (b) Schematic top view of the studied Hall-bars.
   (c) X-ray diffraction curves of the system obtained in the opposite directions for a  [026] reflex (for azimuth angles $\varphi=0^{\circ}$ and $180^{\circ}$).
   The HgTe peak positions relative to CdTe allow determining its average strain which is about 60\% (100\% means that the HgTe film has the CdTe lattice constant and the 15\,meV indirect bulk band gap, 0\% means that it is fully relaxed to its own lattice constant and has some meV of the bands overlap \cite{Berchenko1976}).
   } \label{fig: samples}
\end{figure}

In Fig.~\ref{fig: samples}\,(a) we schematically show cross-section view of the system under study.
The 200\,nm HgTe film is sandwiched between thin Cd$_{0.6}$Hg$_{0.4}$Te barriers, a Ti/Au gate has been deposited on the 200\,nm SiO$_2$ insulator grown by a low temperature chemical vapor deposition process.

An approximate thickness of the pseudomorphic growth of a HgTe film on a CdTe substrate with a 0.3\%
larger lattice constant is about 100 -- 150\,nm  (according to \cite{Brune2011} and our experience).
The full 100\%  adoption of the CdTe substrate lattice constant results in a uniaxial strain and opening of an indirect 15\,meV bulk band gap \cite{Dantscher2015}.
To reveal the crystalline perfection of the grown 200\,nm HgTe structure, a two-crystal X-ray diffractometer was used (Cu-K$_\alpha$ radiation, a Ge (400) monochromator).
Shown in Fig.~\ref{fig: samples}\,(c) are the high resolution diffraction curves of a [026] reflex in a symmetrical Bragg geometry.
The data confirm the suggested relaxation, but not full: the structure still adopts the CdTe substrate, but has the lattice constant, which is an average of bulk CdTe and HgTe with the weights of 60\% and 40\%, respectively. As compared with thinner (with thickness below $100$\,nm) HgTe films, the film under study is only 60\% strained \footnote{The HgTe diffraction peaks symmetry in the X-ray spectra proves that the HgTe film is strained uniformly since any inhomogeneity would lead to a noticeable asymmetry of the peaks.}. In turn, the reduced strain should lead to a smaller value of the band gap. The following transport measurements show nearly a zero indirect band gap.

Several Hall-bar samples were prepared from one wafer showing similar results. For the consistency of magnetotransport and capacitance responses,  all the presented data are from one sample.
In Fig.~\ref{fig: samples}\,(b) we show a schematic top view of the studied Hall-bars with a 50\,$\mu$m current channel and equal to 100 and 250\,$\mu$m distances between potential probes. 
In order to enhance the signal-to-noise ratio during the capacitance measurements, the gate covered not only the central part of the Hall-bar, but a significant part of the contact leads as well. The total area of the gated region was around 1.1\,mm$^2$.
Transport measurements were performed using a standard lock-in technique with a driving current in the range of 10$^{-7}$ -- 10$^{-8}$\,A in a perpendicular magnetic field $B$ up to 12\,T at the temperature 0.2\,K. A typical frequency for transport measurements was 12\,Hz that was decreased to 2\,Hz in the case of high-magnetic fields. For the capacitance measurements we mixed the dc bias $V_\text{g}$ with a small ac voltage $V_\text{ac}$ and measured the ac current flowing across our device phase sensitively. The total capacitance measured in such way between the metallic top gate and a 2D electron system (2DES) depends, besides the geometric capacitance, on the quantum capacitance $e^2D$, connected in series and reflecting the finite density of states $D$ of 2DES \cite{Smith1985, Kozlov2016}; $e$ is the elementary charge, $D$ is the thermodynamic density of states.
The ac frequency for capacitance measurements was in the range of 2 -- 680\,Hz.
The frequency independence of measured resistance and capacitance $C$ was controlled excluding both the existence of leakage currents and resistive effects. The parasitic capacitance of our set up was about 40\,pF.

\section{Results and discussion}\label{Sec: Results and discussion}

The detailed analysis of the magnetotransport and capacitance data reveals the following main idea: the spin non-degenerate surface states, originated from the HgTe inverted band structure , persist in the 200\,nm partially relaxed film and act in a very similar way as in strained films with bulk gaps.
Like it is for 80\,nm HgTe films \cite{Kozlov2014,Kozlov2016}, the surface carriers of the system are high mobility electrons, while in the bulk one can observe both electrons and holes depending on the Fermi level position.

In the inset of Fig.~\ref{fig: cl_tr}\,(a) we show a suggested schematic band diagram of the system at zero gate voltage. In terms of a band structure, the partially relaxed 200\,nm film behaves like an intermediate between a 80\,nm film \cite{Kozlov2014} and bulk HgTe \cite{Berchenko1976}.
While the conduction band has its minimum at the $\Gamma$ point,
the top of the valence band is situated aside of the Brillouin zone center with an indirect energy bulk gap of around 3\,meV.
The band inversion symmetry of the system results in the formation of the topological surface states \cite{Ando2013} (shown in red in Fig.~\ref{fig: cl_tr}\,(a)) with linear-like dispersion and the Dirac point located deep in the valence band \cite{Brune2011}.

\subsection{Classical transport and Drude fitting}\label{SubSec: Classic transport}

\begin{figure}[h]
	\includegraphics[width=1\columnwidth]{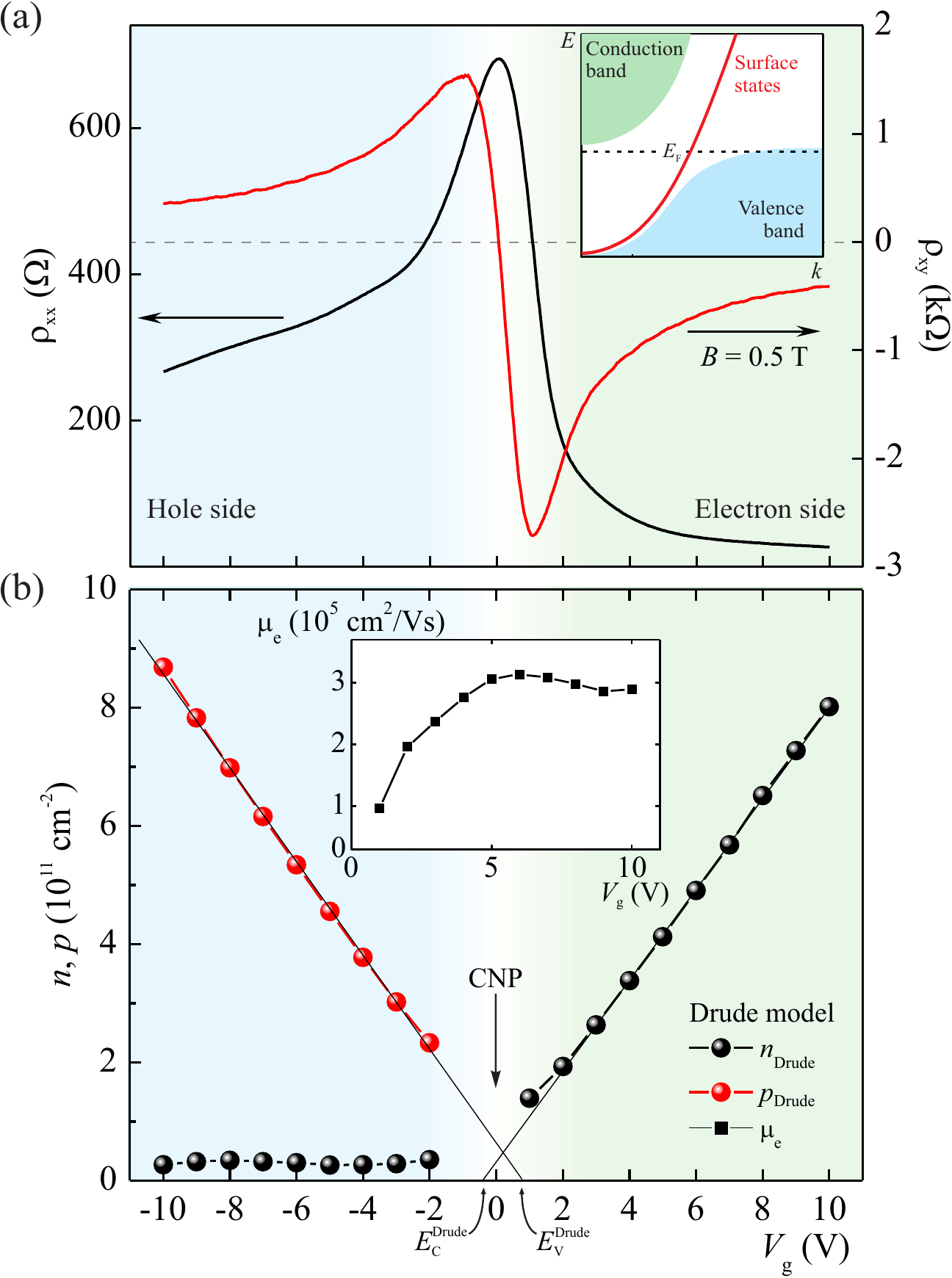}
	\caption{
		(a) The gate dependences of longitudinal resistivity $\rho_\text{xx}$ (measured at zero magnetic field, black line) and Hall resistance $\rho_\text{xy}$ (at magnetic field $B=0.5$\,T, red line). $\rho_\text{xx}$ exhibits its maximum, while $\rho_\text{xy}$ changes the sign in the vicinity of a charge neutrality point (CNP) of the system.
		\textit{Inset} -- a schematic band diagram of the system at zero gate voltage.
		(b) The gate voltage dependences of the total electron $n_\text{Drude}$ (black spheres) and hole $p_\text{Drude}$ (red spheres) densities obtained from the two-component Drude model fitting of the $\rho_\text{xx}(B)$ and $\rho_\text{xy}(B)$ traces.
		The linear extrapolation of $p_\text{Drude}$ and $n_\text{Drude}$ to zero values results in the highlighting of two points on the $V_\text{g}$-axis for holes and electrons, respectively, marked as $E_\text{V}^\text{Drude}$ at $V_\text{g} \approx 0.8$\,V and $E_\text{C}^\text{Drude}$ at $V_\text{g} \approx -0.3$\,V. The first point corresponds to the state of the system when the Fermi level touches the top of the valence band. However, the second point has only the virtual meaning without any well-defined state of the system (see text for details).
		\textit{Inset} -- the gate dependence of the average electron mobility $\mu_\text{e}$ at positive gate voltages.
	} \label{fig: cl_tr}
\end{figure}

Shown in Fig.~\ref{fig: cl_tr}\,(a) are the typical resistivity $\rho_\text{xx}$ at magnetic field $B = 0$ and Hall resistance $\rho_\text{xy}$ at $B = 0.5$\,T as a functions of gate voltage
$V_\text{g}$ at $T=0.2$\,K. The $\rho_\text{xx}$ trace exhibits a maximum near $V_\text{g} =1$\,V and is asymmetric with respect to $V_\text{g}$: the resistivity on the left side of the maximum is significantly higher than that on
the right side. 
In the vicinity of the $V_\text{g} = 0$\,V point $\rho_\text{xy}$ changes its sign. 
The obtained behavior is very similar to the observed one in thinner HgTe films \cite{Kozlov2014} and in line with the expectations that the gate voltage changes the position of the Fermi level from the valence to conduction band. 
On the right side of the peak, where $\rho_\text{xy}$ is negative, the Fermi level goes to the conduction band and the carriers are the high-mobility surface electrons and moderate-mobility bulk ones. 
In this region $\rho_\text{xx}$ reaches its minimal values. 
At negative gate voltages, the Fermi level goes to the valence band, where the bulk holes and the surface electrons coexist resulting in the non-linear sign-variable Hall effect similar to what was observed in 80\,nm films \cite{Kozlov2014} (not shown). 
An almost one order difference in $\rho_\text{xx}$ values between left and right sides of the resistivity peak indicates a significant difference in electron and hole mobilities.

In order to obtain the values of electron and hole densities and mobilities ($n_\text{Drude}$ and $p_\text{Drude}$, $\mu_\text{e}$, and $\mu_\text{p}$, respectively) we used the two-component Drude model fitting of both $\rho_\text{xx}(B)$ and $\rho_\text{xy}(B)$ dependences at fixed gate voltages, as it had been carried out in our former works \cite{Kvon2008, Kozlov2014}. 
One should note that practically there are more than one kind of electrons  in the system, namely: two kinds of the surface electrons (from the top and bottom surfaces) and the bulk carriers. However, the tolerance of the Drude model does not allow distinguishing between them. 
Therefore, the values of $n_\text{Drude}$ and $\mu_\text{e}(V_\text{g})$, obtained from the fitting,  reveal the total electron density and the average electron mobility.
The obtained by this manner $n_\text{Drude}(V_\text{g})$ and $p_\text{Drude}(V_\text{g})$ density dependences are shown in Fig.~\ref{fig: cl_tr}\,(b). 
The mobility dependence $\mu_\text{e}(V_\text{g})$ for the region $V_\text{g}>0$ is shown in the inset.
At a negative gate voltage side, where the bulk holes and  surface electrons coexist, the performed fitting provides both densities, but only the hole mobility, while the electron mobility remains uncertain. The hole mobility is proved to be gate independent and has a value of about 10$^4$\,cm$^2$/Vs (not shown).

The density dependences obtained from the Drude fitting can be interpreted in the following way.
At zero gate voltage the Fermi level is located near the top of the valence band, where the bulk holes coexist with nearly the same number of electrons, presumably the surface ones. By applying small positive $V_\text{g}$ the total electron density $n_\text{Drude}(V_\text{g})$ increases, though not with the full filling rate because of the bulk hole presence. At $V_\text{g} = E_\text{V}^\text{Drude}$ (see the caption of Fig.~\ref{fig: cl_tr}\,(b)) the Fermi level touches the top of the valence band.
At $V_\text{g} > E_\text{V}^\text{Drude}$  the bulk holes disappear, while the electron density $n_\text{Drude}$ increases linearly with $V_\text{g}$ with the full filling rate.
On the contrary, by applying negative $V_\text{g}$ one mainly increases $p_\text{Drude}$, while $n_\text{Drude}(V_\text{g})$ stays almost constant because of much smaller effective mass values and, hence, those of the density of states. A small increase of $n_\text{Drude}(V_\text{g})$ observed at $V_\text{g} < -5$\,V has no physical sense and, probably, indicates  a fitting uncertainty in this range.
The total filling rate $dn/dV_\text{g}$ describes the change of the total carrier density $(n_\text{Drude} - p_\text{Drude})$ with $V_\text{g}$ and is mainly determined by the insulator layer sequence.
The filling rate obtained from the experiment is found to be $8.0 \times 10^{10}$\,cm$^{-2}$/V, which is in line with electrostatic calculations.

One should note that the Drude fitting gives only the total electron density value and not partial ones.
Therefore, the Drude fitting does not reveal the actual position of the bottom of the conduction band. 
The obtained from a linear extrapolation the $E_\text{C}^\text{Drude}$ point on the $V_\text{g}$-axis does not correspond to any well-defined state of the system and has only a virtual sense. Consequently it is impossible to extract the bulk energy gap value from the classical magnetotransport data. More information can be obtained only from an analysis of the quantum oscillations.

\subsection{Quantum Transport and Capacitance Oscillations}\label{SubSec: Capacitance oscillations}
In contrast to the classical transport, the analysis of Shubnikov-de Haas (SdH) oscillations in systems with high enough quantum mobility allows determining the partial densities of all carriers \cite{Fletcher2005, Kozlov2014}. One can expect four groups of carries in the system under study, though, at $V_\text{g} > E_\text{V}^\text{Drude}$ only three of them are left. Each group of carriers is characterized by a density giving its own frequency of the SdH oscillations on a $1/B$ scale. The issue of carriers identification from the resistivity oscillations with three kinds of electrons in a general case looks rather complicated and its solution is too ambiguous.
On the other hand, the issue can be simplified if one complements the resistivity oscillations with the capacitance ones. In case of ordinary 2DES the capacitance oscillations simply reflect the density of states modulation by a magnetic field. However, in the case of an essentially thick film with several groups of carriers that are located at different areas across the film, it turns out that the amplitude of the capacitance oscillations produced by each group is in the inverse relation with a distance from the carriers to the gate \cite{Kozlov2016, Inhofer2017}. Therefore one can expect that the amplitude of the SdH oscillations observed in capacitance induced by the top surface electrons is most enhanced, while the electrons on the bottom surface, as the most distant from the gate, produce the oscillations of a suppressed amplitude. By the comparison of the oscillations in capacitance and transport and their Fourier spectra, measured at the same conditions, one can identify what is what.

\begin{figure}
	\includegraphics[width=0.95\columnwidth]{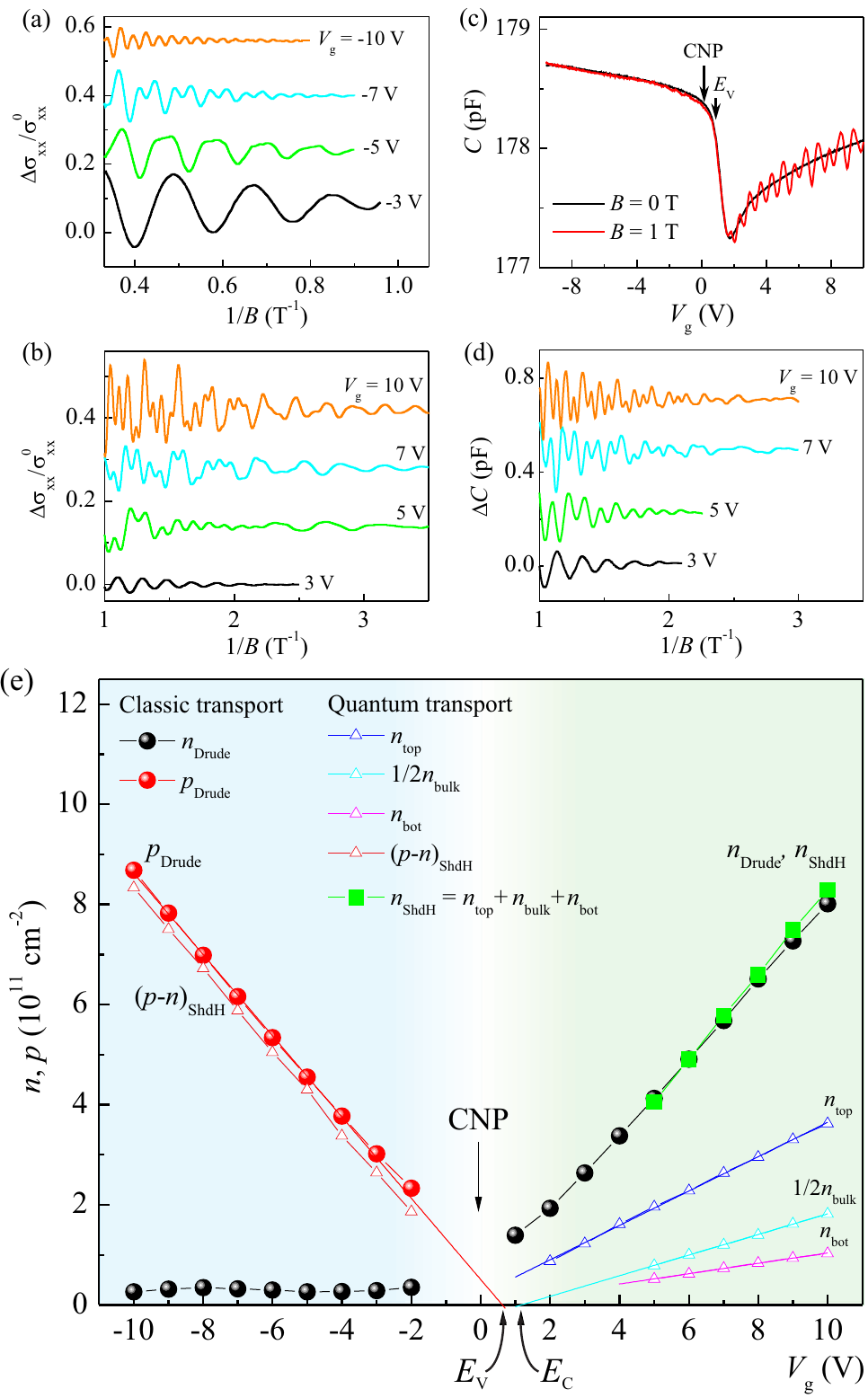}
	\caption{
		(a), (b) The normalized conductivity oscillations $\Delta \sigma_\text{xx} / \sigma_\text{xx}^0 = \bigl(\sigma_\text{xx} - \langle \sigma_\text{xx} \rangle \bigr) / \sigma_\text{xx}^0$ on the $B^{-1}$ scale for the hole and electrons sides, respectively, where $ \langle \sigma_\text{xx} \rangle $ denotes the monotonous part of $\sigma_{xx}$ and $\sigma_{xx}^0 \equiv \sigma_{xx}(B=0)$.
		(c) The gate dependences of the capacitance $C(V_\text{g})$ measured at zero magnetic field black line) and at $B=1$\,T (red line).
		(d) The SdH oscillations in $\Delta C = C - C(B=0)$ for the electron side.
		(e) The combined gate voltage density dependences obtained from both classical transport and SdH oscillations analyses.
		The above presented in Fig.~\ref{fig: cl_tr}\,(b) $n_\text{Drude}$ and $p_\text{Drude}$ have the same color code and are shown in black and red spheres, respectively. From SdH oscillations the following densities were obtained:
		the differential density $(p - n)_\text{SdH}$ for the hole side, and partial densities $n_\text{top}$, $n_\text{bulk}$ and $n_\text{bot}$ for the electron side. 
		The $1/2 \cdot n_\text{bulk}$ is shown for the consistency with a FFT spectral analysis.
		Also the total electron density, obtained as $n_\text{SdH}= n_\text{top} + n_\text{bulk} + n_\text{bot}$ is shown in green squares.
	} \label{fig: Osc}
\end{figure}

In Fig.~\ref{fig: Osc} we show the normalized SdH conductivity oscillations  $\Delta \sigma_\text{xx} / \sigma_\text{xx}^0$
at different gate voltages (panels (a) and (b)) vs a reciprocal magnetic field $B^{-1}$, the gate dependence of measured capacitance $C(V_\text{g})$ at zero and non-zero magnetic field (c) and its oscillations $\Delta C(B^{-1})$ for different $V_\text{g}$ values (d). Each oscillation pattern in the $B^{-1}$ scale was subject to a fast Fourier transform (FFT) analysis (see the Appendix for details).

\begin{figure}
	\includegraphics[width=0.8\columnwidth]{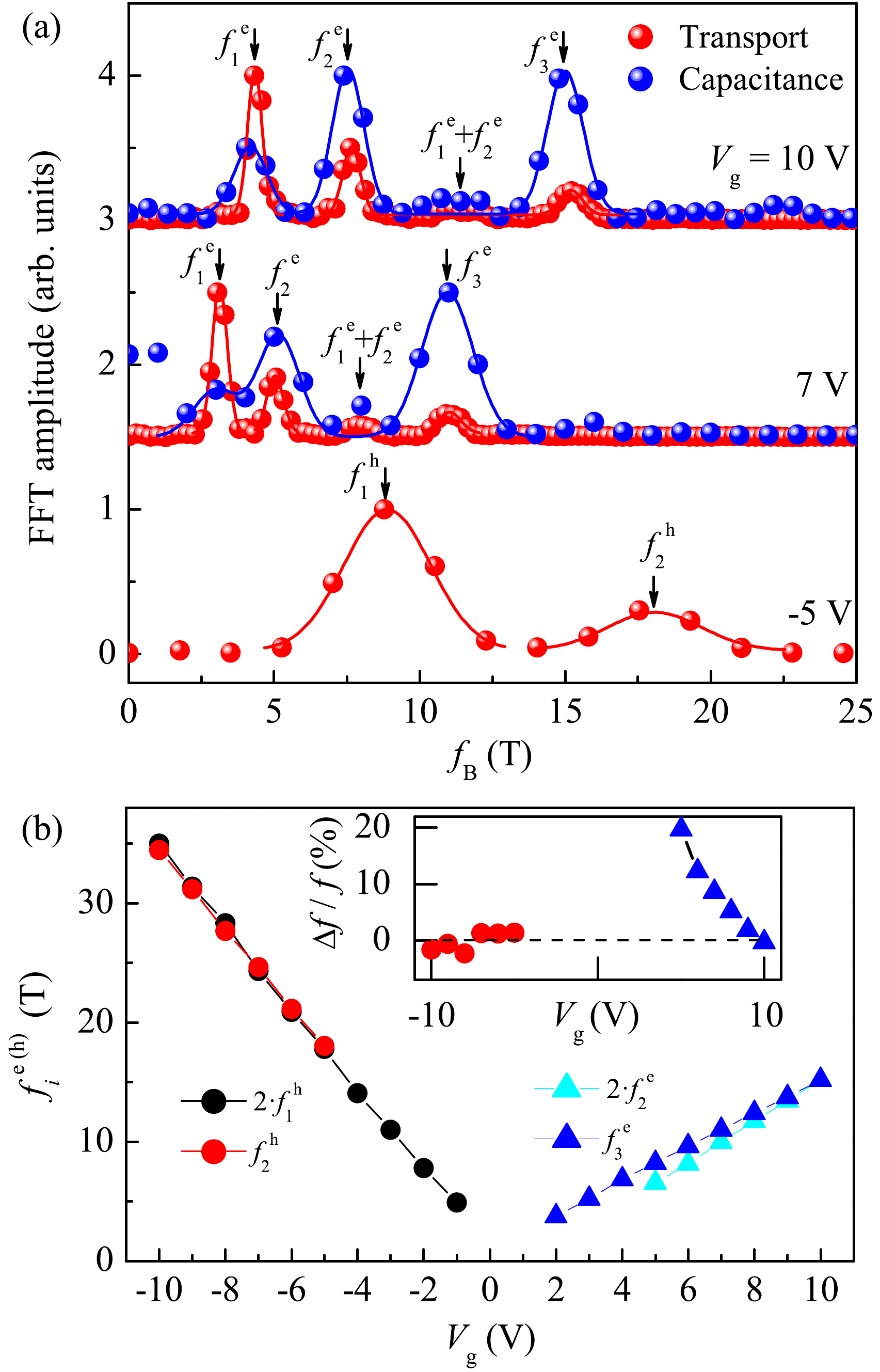}
	\caption{
		(a) The typical Fourier spectra of the conductivity (red spheres) and capacitance (blue spheres) SdH oscillations at different fixed gate voltages. 
		The solid lines correspond to the fitting by Gaussian functions. 
		The vertical arrows point out the central frequency of each Gaussian fit.
		(b) The ratio between positions of different peaks. 
		On the hole side $2\cdot f_1^\text{h}(V_\text{g})$ (black circles) closely follows the $f_2^\text{h}(V_\text{g})$ (red circles) dependence indubitably pointing out their common origin, namely spin-degenerated and spin-resolved  Landau levels of bulk holes, respectively. 
		In contrast, on the electron side the $2\cdot f_2^\text{e}(V_\text{g})$ (cyan triangles) and $f_3^\text{e}(V_\text{g})$ (blue triangles) dependences demonstrate different slopes indicating the different nature of the peaks characterized by these frequencies. 
		\textit{Inset} -- the enhanced difference between hole $( f_2^\text{h} - 2\cdot f_1^\text{h}) / f_2^\text{h}$ (red circles) and electron $(f_3^\text{e} - 2\cdot f_2^\text{e}) / f_3^\text{e}$ (blue triangles) traces at the same values of $V_\text{g}$.
	} \label{fig: AppOsc}
\end{figure}

On the hole side of the gate voltages the SdH oscillations (Fig.~\ref{fig: Osc} (a)) demonstrate the simplest pattern, and their Fourier spectra demonstrate two well-defined peaks (see Fig.~\ref{fig: AppOsc} (a)). 
The behavior is typical for 2D systems, where high-mobility and low-density electrons coexist with low-mobility and high-density holes. 
Because of their low density, the electrons do not produce any well-resolved peaks on the Fourier spectrum. On the contrary, the holes induce the SdH oscillations accompanied with two well-resolved peaks on the FFT spectrum of frequencies $f_1^\text{h}$ and $f_2^\text{h}$. For every gate voltage these frequencies are governed by the relation $2\cdot f_1^\text{h} = f_2^\text{h}$ (see Fig.~\ref{fig: AppOsc} (b)) and, therefore, correspond to a formation of spin-degenerated and spin-resolved bulk hole Landau levels, respectively. However, the exact frequency values correspond to the differential density $(p-n)$ and not to the hole one \cite{Mendez1985, Kozlov2014}. The gate voltage dependence of $(p-n)_\text{SdH}(V_\text{g}) = 2 e/h \cdot f_1^\text{h}$ is shown as red triangles in Fig~\ref{fig: Osc} (e); $h$ is the Planck constant. The obtained data are in full agreement with the Drude fitting, since it is clearly seen that the $(p-n)_\text{SdH}(V_\text{g})$ dependence accurately follows $p_\text{Drude}(V_\text{g})$ with a shift of around $n_\text{Drude}(V_\text{g})$, i.e. $p_\text{Drude}(V_\text{g}) - n_\text{Drude}(V_\text{g}) = (p-n)_\text{SdH}(V_\text{g})$.

On the positive gate voltage side the SdH oscillations demonstrate a much more complicated structure both in the conductivity and capacitance (see Fig.~\ref{fig: Osc} (b) and (d)). Since the high electron mobility, the first oscillations appear at the magnetic fields as low as 0.25\,T, while at $B>1$\,T the transition to the quantum Hall state begins (not shown). Because of our focus on the SdH oscillations, we limited the magnetic field range to 1\,T.
For the traces measured at $V_\text{g} \geq 5$\,V the Fourier spectra demonstrate three independent well-resolved peaks (see Fig.~\ref{fig: AppOsc} (a)), characterized by the frequencies of $f_1^\text{e}$, $f_2^\text{e}$ and $f_3^\text{e}$. 
Also the additional peaks with the frequencies of $f_\text{i}^\text{e} \pm f_\text{j}^\text{e}$ and of much smaller amplitudes are seen. 
The comparison of the conductivity and capacitance Fourier spectra reveals that the relative peak heights significantly differ between two techniques: the peak with frequency of $f_3^\text{e}$ is strongly enhanced in the capacitance spectra (compare to its amplitude in the corresponding transport spectra), and it points out that it stems from the top surface electrons with the density of $n_\text{top} = e/h \cdot f_3^\text{e}$ (see Fig.~\ref{fig: Osc} (e)).
On the contrary, the peak of $f_1^\text{e}$ frequency is strongly suppressed in the capacitance response indicating that it originates from the bottom surface electrons with the density of $n_\text{bot} = e/h \cdot f_1^\text{e}$. Then, the  peak with the frequency of $f_2$ should be associated with the bulk electrons. Since the bulk electrons are expected to be spin-degenerated, then their density should be governed by the relation $n_\text{bulk} = 2e/h \cdot f_2^\text{e}$.
One should note that the spin degeneracy of the bulk carriers should be removed in strong magnetic fields. Indeed, the SdH oscillations splittings were observed in magnetic fields up to 5-12\,T, but the corresponding Fourier spectra did not demonstrate a well-resolved peak with the frequency of $2\cdot f_2^\text{e}$ probably because of the proximity to the $f_3^\text{e}$ peak.
At $V_\text{g} < 5$\,V the peaks with frequencies of $f_1^\text{e}$ and $f_2^\text{e}$ become hardly visible, while the $f_3^\text{e}$ peak persists.

An independent proof of supposed carriers identification can  be obtained by a comparison of the densities, resulted from the SdH oscillations, with the total density obtained from the Drude fits. An almost perfect coincidence of $n_\text{Drude}(V_\text{g})$ with $n_\text{top} + n_\text{bot} + n_\text{bulk}$ was obtained (see Fig.~\ref{fig: Osc} (e)). All other possible correspondences
of the Fourier peaks with the groups of carriers were checked, but a much worse agreement in the relation $n_\text{Drude} = e/h\cdot(\sum g_\text{i} f_\text{i}^\text{e})$ was obtained, where $g_\text{i}=1,2$ are possible spin degeneracy values. An additional check of that $f_3^\text{e} \neq 2 \cdot f_2^\text{e}$ was performed proving that the peaks with $f_2^\text{e}$ and $f_3^\text{e}$ frequencies have  different origins. Indeed, as one can easily see in Fig.~\ref{fig: AppOsc} (b), $f_3^\text{e}(V_\text{g})$ and $2\cdot f_2^\text{e}(V_\text{g})$ dependences have different slopes with an interception around $V_\text{g} = 10$\,V indicating their distinct origin.

Conclusively, the combined analysis of the capacitance and magnetotransport data allows distinguishing and identifying three groups of electrons. Additional information about the band structure can be obtained by a linear extrapolation of the $n_\text{bulk}(V_\text{g})$ dependence (Fig.~\ref{fig: Osc} (e)). Its cross-section with the horizontal axis occurs at $V_\text{g}=1.2$\,V and marks $E_\text{C}$ as a conduction band bottom. The valence band top $E_\text{V}$ was correctly determined by the classical transport analysis and is located at $V_\text{g}=0.8$\,V. The difference between $E_\text{C}$ and $E_\text{V}$ defines the bulk energy gap. The obtained value of $0.4$\,V is 5 times smaller than in the strained HgTe films \cite{Kozlov2014, Kozlov2016}. Therefore, one can estimate that the energy gap value in the partially relaxed HgTe film is about 2-3\,meV. One should note that the obtained indirect energy gap value has the same order as the characteristic disorder in the system.

\section{Conclusion}
In conclusion, by combining the transport measurements with the capacitance spectroscopy, we performed the analysis of the 200\,nm partially relaxed HgTe film.
The Drude fit of the classical magnetotransport measurements reveals the ambipolar electron-hole transport of high electron mobility. The detailed analysis of the Shubnikov-de Haas oscillations, both in conductivity and capacitance, allowed distinguishing three electron groups, identified as electrons on the top and bottom surfaces and the bulk electrons. The indirect bulk energy gap value was found to be 3\,meV. 
It has been established that the significant reduction of the bulk gap does not affect the main properties of the surface states that are found to be spin non-degenerate. 
The results of the work also confirm that the topological surface states are robust to the existence of trivial carriers in a quasi-three-dimensional system.
The presented techniques allow investigations of other 3D TIs regardless of a bulk gap or conductance existence.

\begin{acknowledgments}
	
	This work supported by RFBR Grants No. 18-32-00138 (transport results) and 18-42-543013 (together with the Government of the Novosibirsk Region of the Russian Federation). D.K. was supported by Russian President grant MK-3603.2017.2.
	
\end{acknowledgments}

\appendix*

\section{SdH-oscillations analysis}\label{SubSec: bulk}
The analysis of Shubnikov-de Haas oscillations allows determining the carriers density at a rather high precision, but in the case of several types of carriers the procedure should be performed carefully. Here we describe how the oscillations were processed.

At the first step we measured the wide range magnetic field dependences of the longitudinal $\rho_\text{xx}$ and perpendicular $\rho_\text{xy}$ components of a resistivity tensor. Then we calculated the experimental conductivity trace $\sigma_\text{xx}(B^{-1})$, with which it is better to work because of a smaller monotonous part, but which still has to be removed. The monotonous part, denoted as $ \langle \sigma_\text{xx} \rangle $, was obtained by the smoothing of every $\sigma_\text{xx}(B^{-1})$ trace with an averaging period higher than a characteristic oscillations period. Finally, the oscillatory conductivity part was normalized using the formula $\Delta \sigma_\text{xx}/\sigma_\text{xx}^0 = (\sigma_\text{xx}(B^{-1}) - \langle \sigma_\text{xx}(B^{-1}) \rangle )/\sigma_\text{xx}(B=0)$ and it is shown in
Fig.~\ref{fig: Osc}\,(a), (b). The last step before the Fourier transform was the magnetic field range shrinking, because small fields do not consist of any useful information while at high fields the transition to the quantum Hall state begins. On the hole side the optimal range was found to be from 1.1\,T to 3\,T, while on the electron side the range was limited from 0.2-0.4 to 1\,T depending on the gate voltage. The maximum oscillations amplitude was about 10-15\%.

The procedure was rather similar for the capacitance traces $C(B)$. The only difference was the absence of the monotonous part so $\Delta C(B^{-1}) = C(B^{-1}) - C(B=0)$. For the electron side the magnetic field range was limited from 0.3-0.4 to 1\,T depending on the gate voltage. For the hole side the capacitance measurements were not performed.

Finally, every oscillatory trace was subject to the fast Fourier transform. The typical obtained spectra are shown in Fig.\ref{fig: AppOsc}\,(a). Each spectrum demonstrates a set of peaks which are characterized by frequencies $f_\text{i}$. The best known way to determine the precise $f_\text{i}$ value, especially in case of closely spaced peaks, is fitting each peak with any dome-like dependence. We used Gaussian functions $A_\text{i}\exp(-\frac{(f-f_\text{i})^2}{2\Delta f^2})$  with a central frequency $f_\text{i}$ for the fitting of each peak. The obtained fitting is shown in Fig.\ref{fig: AppOsc}\,(a) by solid lines. The corresponding central frequencies are marked by vertical arrows.

\bibliography{library}
\end{document}